\documentclass[times,tighten,twocolumn,xenglish]{aastex631}
\usepackage{lipsum}
\usepackage{physics}
\usepackage{multirow}
\usepackage{rotating}
\usepackage{graphicx}
\usepackage{subfigure}
\usepackage{color}
\usepackage{amsfonts}
\usepackage{mathrsfs}
\usepackage{babel}
\usepackage{amssymb}
\usepackage{wasysym}
\usepackage{verbatim}
\usepackage{multirow}
\usepackage{hyperref}
\usepackage{float}
\usepackage{ulem}

\usepackage{paralist}

\begin{document}
\title{The first identification of  Lyman $\alpha$ Changing-look Quasars at
 high-redshift in DESI}

\author[0000-0001-9457-0589]{Wei-Jian Guo}
\affiliation{Key Laboratory of Optical Astronomy, National Astronomical Observatories, Chinese Academy of Sciences, Beijing 100012, China\\ Email:\href{mailto:guowj@bao.ac.cn}{guowj@bao.ac.cn}}

\author[0000-0003-0230-6436]{Zhiwei Pan}
\affiliation{Kavli Institute for Astronomy and Astrophysics at Peking University, PKU, 5 Yiheyuan Road, Haidian District, Beijing 100871, P.R. China}

\author[0000-0002-2949-2155]{Małgorzata Siudek}
\affiliation{Instituto Astrofisica de Canarias, Av. Via Lactea s/n, E38205 La Laguna, Spain}
\affiliation{Institute of Space Sciences, ICE-CSIC, Campus UAB, Carrer de Can Magrans s/n, 08913 BELlaterra, Barcelona, Spain}

\author{Jessica Nicole Aguilar}
\affiliation{Lawrence Berkeley National Laboratory, 1 Cyclotron Road, Berkeley, CA 94720, USA}
\author[0000-0001-6098-7247]{Steven Ahlen}
\affiliation{Physics Dept., Boston University, 590 Commonwealth Avenue, Boston, MA 02215, USA}
\author[0000-0001-9712-0006]{Davide Bianchi}
\affiliation{Dipartimento di Fisica ``Aldo Pontremoli'', Universit\`a degli Studi di Milano, Via Celoria 16, I-20133 Milano, Italy}
\author{David Brooks}
\affiliation{Department of Physics \& Astronomy, University College London, Gower Street, London, WC1E 6BT, UK}
\author{Todd Claybaugh}
\affiliation{Lawrence Berkeley National Laboratory, 1 Cyclotron Road, Berkeley, CA 94720, USA}
\author{Kyle Dawson}
\affiliation{Department of Physics and Astronomy, The University of Utah, 115 South 1400 East, Salt Lake City, UT 84112, USA}
\author[0000-0002-1769-1640]{Axel  de la Macorra}
\affiliation{Instituto de F\'{\i}sica, Universidad Nacional Aut\'{o}noma de M\'{e}xico,  Cd. de M\'{e}xico  C.P. 04510,  M\'{e}xico}
\author{Peter Doel}
\affiliation{Department of Physics \& Astronomy, University College London, Gower Street, London, WC1E 6BT, UK}
\author[0000-0003-2371-3356]{Kevin Fanning}
\affiliation{Kavli Institute for Particle Astrophysics and Cosmology, Stanford University, Menlo Park, CA 94305, USA}
\affiliation{SLAC National Accelerator Laboratory, Menlo Park, CA 94305, USA}
\author[0000-0002-2890-3725]{Jaime E. Forero-Romero}
\affiliation{Departamento de F\'isica, Universidad de los Andes, Cra. 1 No. 18A-10, Edificio Ip, CP 111711, Bogot\'a, Colombia}
\affiliation{Observatorio Astron\'omico, Universidad de los Andes, Cra. 1 No. 18A-10, Edificio H, CP 111711 Bogot\'a, Colombia}
\author{Enrique Gaztañaga}
\affiliation{Institut d'Estudis Espacials de Catalunya (IEEC), 08034 Barcelona, Spain}
\affiliation{Institute of Cosmology and Gravitation, University of Portsmouth, Dennis Sciama Building, Portsmouth, PO1 3FX, UK}
\affiliation{Institute of Space Sciences, ICE-CSIC, Campus UAB, Carrer de Can Magrans s/n, 08913 Bellaterra, Barcelona, Spain}
\author[0000-0003-3142-233X]{Satya  Gontcho A Gontcho}
\affiliation{Lawrence Berkeley National Laboratory, 1 Cyclotron Road, Berkeley, CA 94720, USA}
\author{Klaus Honscheid}
\affiliation{Center for Cosmology and AstroParticle Physics, The Ohio State University, 191 West Woodruff Avenue, Columbus, OH 43210, USA}
\affiliation{Department of Physics, The Ohio State University, 191 West Woodruff Avenue, Columbus, OH 43210, USA}
\affiliation{The Ohio State University, Columbus, 43210 OH, USA}
\author{Robert Kehoe}
\affiliation{Department of Physics, Southern Methodist University, 3215 Daniel Avenue, Dallas, TX 75275, USA}
\author[0000-0003-3510-7134]{Theodore Kisner}
\affiliation{Lawrence Berkeley National Laboratory, 1 Cyclotron Road, Berkeley, CA 94720, USA}
\author{Andrew Lambert}
\affiliation{Lawrence Berkeley National Laboratory, 1 Cyclotron Road, Berkeley, CA 94720, USA}
\author[0000-0003-1838-8528]{Martin Landriau}
\affiliation{Lawrence Berkeley National Laboratory, 1 Cyclotron Road, Berkeley, CA 94720, USA}
\author[0000-0001-7178-8868]{Laurent Le Guillou}
\affiliation{Sorbonne Universit\'{e}, CNRS/IN2P3, Laboratoire de Physique Nucl\'{e}aire et de Hautes Energies (LPNHE), FR-75005 Paris, France}
\author[0000-0003-4962-8934]{Marc Manera}
\affiliation{Departament de F\'{i}sica, Serra H\'{u}nter, Universitat Aut\`{o}noma de Barcelona, 08193 Bellaterra (Barcelona), Spain}
\affiliation{Institut de F\'{i}sica dâ€™Altes Energies (IFAE), The Barcelona Institute of Science and Technology, Campus UAB, 08193 Bellaterra Barcelona, Spain}
\author[0000-0002-1125-7384]{Aaron Meisner}
\affiliation{NSF NOIRLab, 950 N. Cherry Ave., Tucson, AZ 85719, USA}
\author[0000-0002-2733-4559]{John Moustakas}
\affiliation{Department of Physics and Astronomy, Siena College, 515 Loudon Road, Loudonville, NY 12211, USA}
\author{Andrea  Muñoz-Gutiérrez}
\affiliation{Instituto de F\'{\i}sica, Universidad Nacional Aut\'{o}noma de M\'{e}xico,  Cd. de M\'{e}xico  C.P. 04510,  M\'{e}xico}
\author{Adam Myers}
\affiliation{Department of Physics \& Astronomy, University  of Wyoming, 1000 E. University, Dept.~3905, Laramie, WY 82071, USA}
\author[0000-0001-6590-8122]{Jundan Nie}
\affiliation{National Astronomical Observatories, Chinese Academy of Sciences, A20 Datun Rd., Chaoyang District, Beijing, 100012, P.R. China \\ Email:\href{mailto:zouhu@bao.ac.cn}{zouhu@bao.ac.cn}}
\author[0000-0003-3188-784X]{Nathalie Palanque-Delabrouille}
\affiliation{IRFU, CEA, Universit\'{e} Paris-Saclay, F-91191 Gif-sur-Yvette, France}
\affiliation{Lawrence Berkeley National Laboratory, 1 Cyclotron Road, Berkeley, CA 94720, USA}
\author{Claire Poppett}
\affiliation{Lawrence Berkeley National Laboratory, 1 Cyclotron Road, Berkeley, CA 94720, USA}
\affiliation{Space Sciences Laboratory, University of California, Berkeley, 7 Gauss Way, Berkeley, CA  94720, USA}
\affiliation{University of California, Berkeley, 110 Sproul Hall \#5800 Berkeley, CA 94720, USA}
\author[0000-0001-7145-8674]{Francisco Prada}
\affiliation{Instituto de Astrof\'{i}sica de Andaluc\'{i}a (CSIC), Glorieta de la Astronom\'{i}a, s/n, E-18008 Granada, Spain}
\author[0000-0001-5589-7116]{Mehdi Rezaie}
\affiliation{Department of Physics, Kansas State University, 116 Cardwell Hall, Manhattan, KS 66506, USA}
\author{Graziano Rossi}
\affiliation{Department of Physics and Astronomy, Sejong University, Seoul, 143-747, Korea}
\author[0000-0002-9646-8198]{Eusebio Sanchez}
\affiliation{CIEMAT, Avenida Complutense 40, E-28040 Madrid, Spain}
\author{Michael Schubnell}
\affiliation{Department of Physics, University of Michigan, Ann Arbor, MI 48109, USA}
\affiliation{University of Michigan, Ann Arbor, MI 48109, USA}
\author[0000-0002-6588-3508]{Hee-Jong Seo}
\affiliation{Department of Physics \& Astronomy, Ohio University, Athens, OH 45701, USA}
\author[0000-0002-3461-0320]{Joseph Harry Silber}
\affiliation{Lawrence Berkeley National Laboratory, 1 Cyclotron Road, Berkeley, CA 94720, USA}
\author{David Sprayberry}
\affiliation{NSF NOIRLab, 950 N. Cherry Ave., Tucson, AZ 85719, USA}
\author[0000-0003-1704-0781]{Gregory Tarlé}
\affiliation{University of Michigan, Ann Arbor, MI 48109, USA}
\author{Benjamin Alan Weaver}
\affiliation{NSF NOIRLab, 950 N. Cherry Ave., Tucson, AZ 85719, USA}
\author[0000-0002-4135-0977]{Zhimin Zhou}
\affiliation{National Astronomical Observatories, Chinese Academy of Sciences, A20 Datun Rd., Chaoyang District, Beijing, 100012, P.R. China \\ Email:\href{mailto:zouhu@bao.ac.cn}{zouhu@bao.ac.cn}}
\author[0000-0002-6684-3997]{Hu Zou}
\affiliation{National Astronomical Observatories, Chinese Academy of Sciences, A20 Datun Rd., Chaoyang District, Beijing, 100012, P.R. China \\ Email:\href{mailto:zouhu@bao.ac.cn}{zouhu@bao.ac.cn}}

\correspondingauthor{Wei-Jian Guo}
\email{guowj@bao.ac.cn}

% add snr discussion
% add X ray data if need 

\begin{abstract}

We present two cases of Ly$\alpha$ changing-look (CL) quasars (J1306 and J1512) along with two additional candidates (J1511 and J1602), all discovered serendipitously at  $z >2$ through the Dark Energy Spectroscopic Instrument (DESI) and the Sloan Digital Sky Survey (SDSS).   It is the first time to capture CL events in Ly$\alpha$ at high redshift, which is crucial for understanding underlying mechanisms driving the CL phenomenon and the evolution of high-redshift quasars and galaxies.  The variability of all four sources is confirmed by the significant change of amplitude in the  $r$  band  ($|r_{\rm DESI}-r_{\rm SDSS}| >0.5 \ \rm mag$).  We find that the accretion rate in the dim state for these CL objects corresponds to a relatively low value ($\mathscr{\dot M} \approx 2\times10^{-3}$), which suggests that the inner region of the accretion disk might be in transition between the Advection Dominated Accretion Flow ($\mathscr{\dot M}<10^{-3}\sim 10^{-2}$) and the canonical accretion disk (optically thick, geometrically thin). However, unlike in C {\sc iv} CL quasars in which broad Ly$\alpha$ remained,  the broad C {\sc iv} may still persist after a CL event occurs in Ly$\alpha$, making the physical origin of the CL and ionization mechanism event more puzzling and interesting.

\end{abstract}
\vspace{-8mm}
\keywords{Accretion (14); Active galaxies (17); Quasars(1319); Supermassive black holes (1663);}

\section{Introduction}

Changing-look (CL) active galactic nuclei (AGN) or quasars exhibit significant variations in their broad emission lines, continuum flux, and absorption features over relatively short timescales, typically ranging from months to years \citep{denney2014, MacLeod2016}. At $z<0.75$, extreme cases of CL-AGNs can transition from Type 1 AGNs with broad emission lines (BELs) to Type 2 AGNs without BELs, or vice versa \citep{LaMassa2015, Runnoe2016, Yang2018, Yan2019, Graham2020}. CL phenomena challenge the orientation-based unification model and offer new opportunities for exploring the structure and dynamics of the accretion disk and the broad line region (BLR, \citealt{Antonucci1993, Wang2007, Ho2008}).

Current research on CL-AGNs primarily focuses on variations in $\rm H\alpha$, $\rm H\beta$, and Mg {\sc ii} \citep{guo2023}. However, the CL properties of UV emission lines, such as Ly$\alpha$ and C {\sc iv}, which originate from the innermost BLR radii \citep{Collin1988} and are more sensitive to changes in the accretion rate, have attracted less attention. \cite{Ross2020} reported three C {\sc iv} CL quasars through repeated SDSS observations, while \cite{Guo2020} identified several UV broad emission line CL events in extreme variability quasars. These high-redshift CL quasars retained broad Ly$\alpha$ components in their dim states, maintaining the quasar classification.  

The Ly$\alpha$ emission line originates in the innermost BLR, where clouds may affect material exchange with the accretion disk. \cite{Kriss2019} performed reverberation mapping of the Ly$\alpha$ line in NGC 5548, revealing a time lag of 5.1 light-days. In contrast, the time lag for the line wings was only 2 days, closely matching the estimated size of the optically emitting component of the accretion disk (approximately 1.56 light-days in the V-band). This finding suggests that the Ly$\alpha$ radiation region can be influenced by small accretion changes in the central ionizing source, revealing the state of the accretion disk in the nuclear area.

In this Letter,  we report two CL quasars and two case candidates for broad Ly$\alpha$ transition, which is never captured at high redshift. The paper is organized as follows. Section \ref{sec_data} describes the data and sample selection based on  DESI and SDSS. Results and Discussions are given in Sections \ref{sec_results}.  Section \ref{sec_conclusion} summarized the paper. Throughout the paper, we use a $\Lambda$CDM cosmology with $H_{0}= \rm{67\ km\  s^{-1}\  Mpc^{-1}}$, $\Omega_{\Lambda}= 0.68$, and $\Omega_{m}= 0.32$ (\citealt{Planck2020}).

\begin{figure*}[t!]
\centering
\vspace{-0.8cm}
\includegraphics[width=0.8\textwidth]{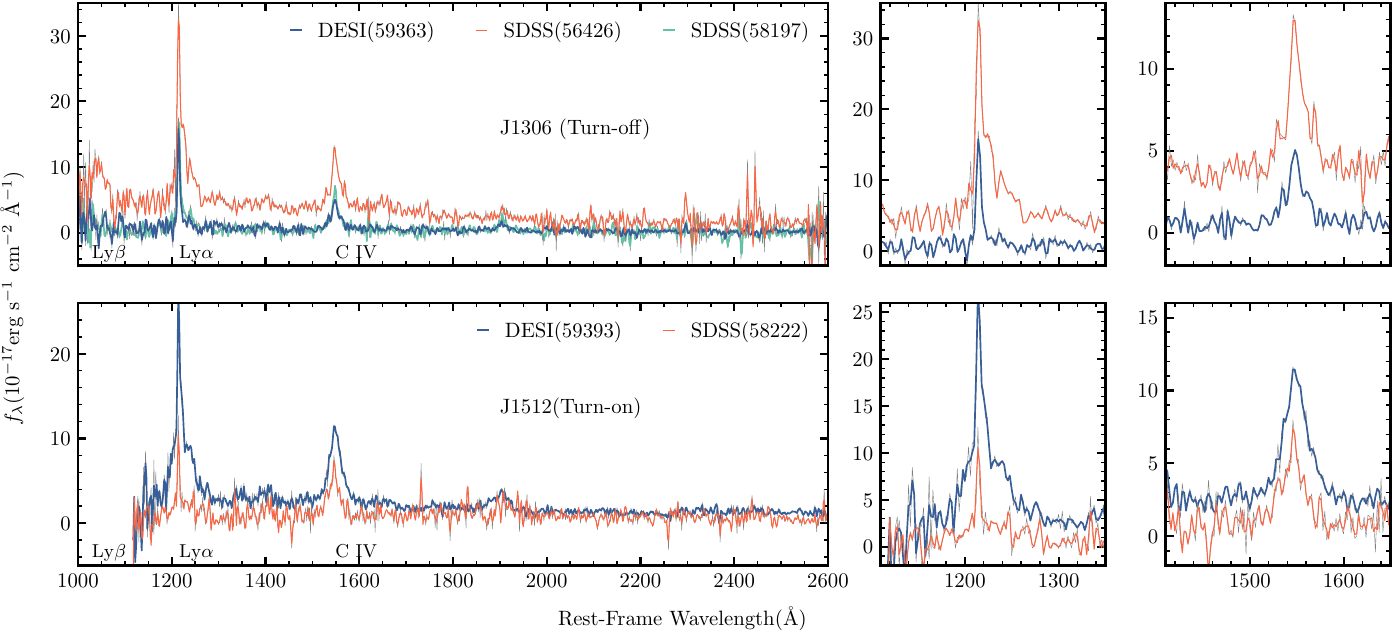}
\includegraphics[width=0.8\textwidth]{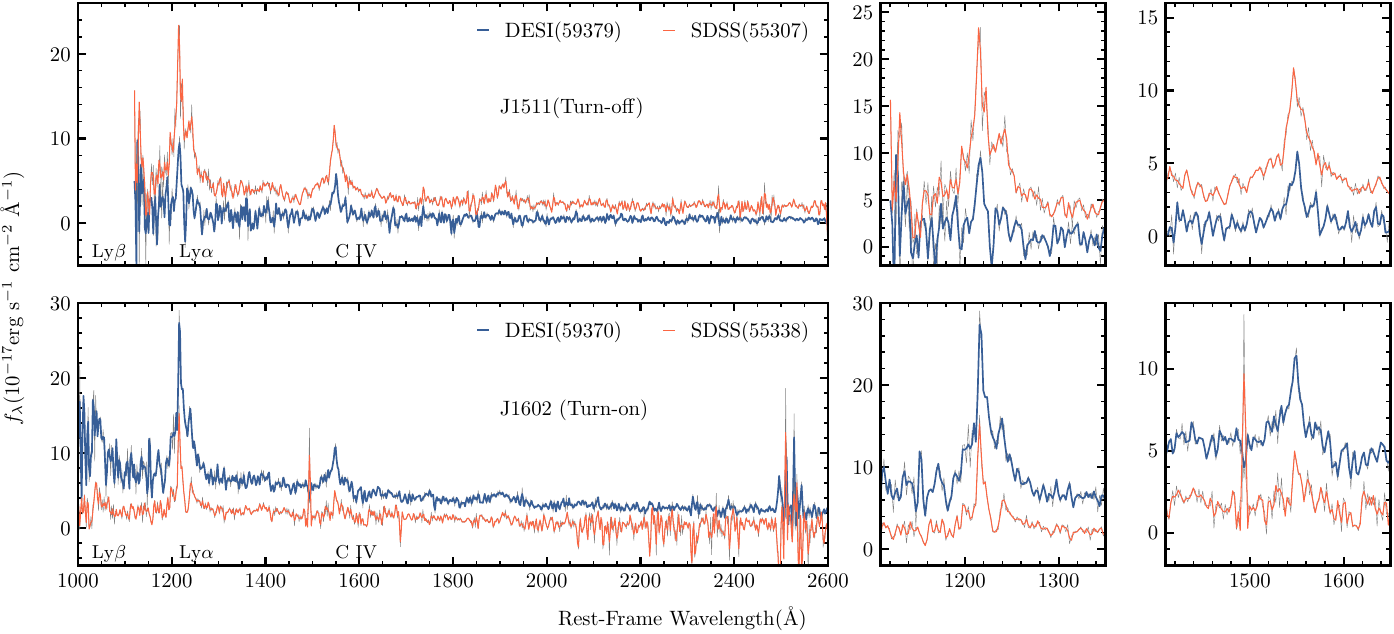}
\caption{The spectra of J1306, J1512, J1511 and J1602 are shown from the top to the bottom panels, with zoomed-in views of the Ly$\alpha$ and C {\sc iv} regions. The orange line represents the SDSS spectra, while the blue line represents the DESI spectra.}
\label{fig_1}
\end{figure*}

\vspace{-4mm}

\section{Data and Selection}
\label{sec_data}

\subsection{Photometry and Spectrum}

The photometric data come from the SDSS imaging survey and the DESI Legacy Surveys\footnote{\url{http://legacysurvey.org/}}, both designed to preselect targets for spectroscopic follow-up surveys \citep{York2000, Gunn2006, Dey2019}. The SDSS imaging survey covers an area of 14,555 deg$^{2}$, with imaging depths in five broad optical bands of approximately $u\approx22.0$ mag, $g\approx22.2$ mag, $r\approx22.2$ mag, $i\approx21.3$ mag, and $z\approx20.5$ mag \citep{Padmanabhan2008, Abazajian2009}. The DESI Legacy Survey has approximate limiting magnitudes in AB magnitude of $g\approx24.0$ mag, $r\approx23.4$ mag, and $z\approx22.5$ mag, covering 9,900 deg$^{2}$ in the North Galactic Cap and 4,400 deg$^{2}$ in the South Galactic Cap \citep{Flaugher2015, Zou2017, Dey2019}.

The  spectra  come from SDSS Data Release 16, spanning approximately 20 years, with a resolution of about $R \approx 1500$ \citep{SDSS_lyke, SDSS_ahumada}. Additionally, the dataset includes spectra from DESI Early Data (``{\em{fuji}}" and ``{\em{guadalupe}}"), with a spectral resolution of approximately $R \approx 3000$ \citep{DESI_Levi, DESI_2016_I,DESI_2016_II, DESI_Abareshi, DESI2022, DESI_Schlafly,DESI_EDR,DESI_SV,DESI_Corrector,DESI_Alexander}.  We mainly use the quasar spectra classified by the DESI ``Redrock'' pipeline \citep{DESI_Guy}.

\begin{deluxetable*}{cccccccccc}
\tablecolumns{10}
\vspace{-0.1cm}
\tabletypesize{\footnotesize}
\tabcaption{\centering The observation details for two Ly$\alpha$ CL quasars (J1306 and J1512) and two candidates (J1511 and J1602).
\label{tab_1}}
\colnumbers
\tablehead{
\colhead{Object} &
\colhead{R.A.} &
\colhead{Dec.} &
\colhead{Redshift} &
 $g$-band & $r$-band & $z$-band & $g$-band & $r$-band & $z$-band \\
 &  (deg)  &  (deg) & &  &
  \colhead{$\rm mag (SDSS)$} &  & &\colhead{$\rm  mag (DESI)$} & }
\startdata
J130643.24+561842.1 & 196.6802  & 56.3117 & 2.638 &  21.78 & 21.55 & 21.42  & 22.63 & 22.74 & 22.71 \\
J151205.89+374627.7 & 228.0245  & 37.7744 & 2.221 & 22.16 & 21.99 & 21.65  &22.50 &  22.56 & 21.83   \\
\hline
J151122.19+291321.8  & 227.8425  & 29.2227 & 2.211 & 21.18 & 21.25 & 21.00   &22.04 &  22.10 &  21.34  \\
J160222.63+174905.5 & 240.5943  & 17.8182 & 2.738 & 22.11 & 21.89 & 21.96  &21.62 &  21.32 &  21.42  
\enddata
\tablecomments{columns: (1) name, (2) right ascension,  (3) declination, (4) redshift, (5)$-$(7) $g$-, $r$-, and $z$-band  magnitude in SDSS image survey, (8)$-$(10)  $g$-, $r$-, and $z$- band  magnitude in DESI image survey.}
\end{deluxetable*}

\begin{deluxetable*}{cccccccccc}
\tablecolumns{10}
\vspace{-0.8cm}
\tabletypesize{\footnotesize}
    \tabcaption{\centering  The physical property measurements of J1306, J1512, J1511 and J1602.
\label{tab_measure}}
\colnumbers
\tablehead{
\colhead{Object} &
\colhead{$\log( L_{1350})$}&
\colhead{$\log( L_{\rm Ly\alpha, all})$}&
\colhead{$\log( L_{\rm Ly\alpha, b})$} &
\colhead{$\log( L_{\text{C\,{\sc iv}}, \rm b})$}&
\colhead{FWHM($\rm Ly\alpha, b$) }&
\colhead{FWHM(C\,{\sc iv}, b) }&
\colhead{$\log(M_{\bullet}/M_{\odot})$} &
\colhead{$\lambda_{\rm Edd} $} &
MJD \\
& $\rm erg\ s^{-1}$ & $\rm erg\ s^{-1}$ 
& $\rm erg\ s^{-1}$ & $\rm erg\ s^{-1}$ 
& $\rm km\ s^{-1}$ &$\rm km\ s^{-1}$ 
& 
}
\startdata
J1306 &  45.54$\pm${0.02}  &  44.48$\pm${0.09} & 44.41$\pm${0.10} & 44.08$\pm${0.08}  &  15472$\pm${4274} & 7667$\pm${1351} & 9.16$\pm${0.12} & 0.072  & 56426 (SDSS)\\
J1306            &  44.55$\pm${0.12}  &  43.98$\pm${0.09} & 43.71$\pm${0.12} &  43.76$\pm${0.11} &  5032$\pm${2002} & 7212$\pm${1984}  &  \nodata & 0.007 & 58197 (SDSS)\\
J1306             &  44.75$\pm${0.11}  &  43.74$\pm${0.06} &  \nodata         & 43.69$\pm${0.13}  & \nodata          & 7224$\pm${2548}  & \nodata & 0.012 & 59363  (DESI)\\
J1512 &  44.64$\pm${0.10}  &  43.29$\pm${0.11} &  \nodata         & 43.81$\pm${0.07} &  \nodata          & 7701$\pm${1559}  & \nodata  & 0.016 &58222 (SDSS)\\
J1512  &  45.15$\pm${0.05}  &  44.34$\pm${0.23} & 44.30$\pm${0.25} & 43.98$\pm${0.05} &  19273$\pm${9461} & 6674$\pm${927}   & 8.92$\pm${0.12} & 0.051  & 59393 (DESI)\\
\hline
J1511 &  45.30$\pm${0.02}  &  44.32$\pm${0.07} & 44.26$\pm${0.08} & 44.03$\pm${0.07}  & 16387$\pm${4273} & 12837$\pm${1596} & 9.56$\pm${0.11} & 0.017 & 55307 (SDSS)\\
J1511            &  44.67$\pm${0.10}  &  43.52$\pm${0.14} &  \nodata         & 43.50$\pm${0.25}  &  \nodata         &10335$\pm${5682}  & \nodata  & 0.004 & 59379  (DESI)\\
J1602 &  45.31$\pm${0.15}  &  43.79$\pm${0.04} &  \nodata         & 43.33$\pm${0.12} &  \nodata          &  \nodata         &  \nodata & 0.020 & 55338 (SDSS)\\
J1602            &  45.74$\pm${0.01}  &  44.52$\pm${0.10} & 44.48$\pm${0.11} & 43.88$\pm${0.15} &  15825$\pm${3325} & 9143$\pm${3109}  & 9.50$\pm${0.30} & 0.053 & 59370 (DESI)\\
\enddata
\tablecomments{columns: (1) name, (2) optical luminosity at 1350~\AA,  (3) Ly$\alpha$ luminosity (narrow line+broad line), (4) Ly$\alpha$ luminosity (broad line), (5) C\,{\sc iv} luminosity (broad line), (6) FWHM of the Ly$\alpha$ broad components, (7) FWHM of the C\,{\sc iv} broad components, (8) black hole mass,  (9)  Eddington ratio  (10) spectral MJD.\\
}
\end{deluxetable*}

\vspace{-12mm}
% \subsection{ Data}

\subsection{Selection}

Following \cite{guo2023}, we cross-matched the DESI and SDSS spectra and applied a redshift cut of $z > 2.1$ to ensure the Ly$\alpha$ line was included in the spectra. We applied the Galactic extinction curve from \cite{Fitzpatrick1999PASP}, assuming $R_{V} = 3.1$, and rebinned the flux and its variance for both the SDSS and DESI spectra with a 2 Å per pixel wavelength.

To quickly select targets with significant Ly$\alpha$ variations, we utilized a significance threshold of $N_{\sigma} > 3$ for the broad emission line (BEL) maximum flux  difference between the dim and bright spectra \citep{MacLeod2019}:
\begin{eqnarray}
\label{eq_N}
N_{\sigma}=(f_{\rm bright }-f_{\rm dim } )/ \sqrt{\sigma^{2}_{\rm bright}+\sigma^{2}_{\rm dim} },
\end{eqnarray}  
where $f$ and $\sigma$ represent the spectral flux and variance, respectively, in units of $\rm erg \ cm^{-2}\ s^{-1} \ Å^{-1}$. Subsequently, we employed spectral decomposition to exclude targets with prominent broad Ly$\alpha$ emission components in the dim state, as illustrated in Appendix \ref{appendix}. Finally, we removed objects with   significant broad Ly$\alpha$ emission lines in the dim state, leading to the identification of two Ly$\alpha$ CL quasars (J130643.24+561842.1 and J151205.89+374627.7) and two candidates (J151122.19 +291321.8 and J160222.63+174905.5).

In \cite{guo2023}, we noted that fiber drop or flux calibration issues may affect the SDSS and DESI spectra. For these four sources, which are fainter than magnitude 21, it is difficult to confirm the CL events through photometric light curves as provided in \cite{guo2023}. However, the consistency between the low-state spectra of J1306 from both SDSS and DESI suggests that the CL event is not due to flux calibration issues. Additionally, the C\,{\sc iv} emission lines in the high and low states of J1512 indicate that the Ly$\alpha$ CL is unlikely to be affected by such issues. On the other hand, we classify J1511 and J1602 as CL candidates for two reasons: (1) the inability to verify flux calibration issues, and (2) the medium full width at half maximum (FWHM) of the narrow Ly$\alpha$ components (1000 km/s $\leq$ FWHM $\leq$ 2000 km/s) in the low state.

\section{Result and Discussion}
\label{sec_results}

Figure~\ref{fig_1} illustrates the spectra of J1306, J1512, J1511 and J1602, where the transition timescales range from 1 to 3.5 years in the rest-frame. For J1306 (a turn-off case), a distinctive raised bump on the blue side of the Ly$\alpha$ line is evident in the SDSS spectrum, made up of Ly$\beta$ and O {\sc vi}, though blended and heavily absorbed by the intergalactic medium (IGM) \citep{Netzer1976, Laor1994, Zheng1995, Bosman2021}. However, in the DESI spectrum, the broad Ly$\beta$/O,{\sc vi} feature has almost disappeared or is overwhelmed by noise as the continuum luminosity decreased. For J1512 (a turn-on case), the broad Ly$\alpha$ emission reappeared in the DESI spectrum, with the profile change being the most significant among the four targets. In the cases of  J1511 (turn-off) and J1602 (turn-on), the CL events in Ly$\alpha$ can be identified through  visual inspection . However, we note that potential artificial flux calibration issues cannot be entirely ruled out, and the spectral signal-to-noise ratio (SNR) in the low state is relatively low.

Table \ref{tab_1} presents observation information for the four quasars, with significant magnitude variations ($|\Delta r| > 0.5 \ \rm mag$) observed between SDSS and DESI photometry.

% Neglecting UV absorption, we attempted to measure the bump in the spectrum of the high states using a Gaussian fit. If assumed to be a symmetrical profile,  the FWHM and shift of the raised bump were up to $15900\pm{3086}$ km/s and $4424\pm{1190}$ km/s, respectively.   Such measurements are overestimated since this raised bump is a composite of the broad Ly$\beta$+O {\sc vi} lines and larger than the Ly$\alpha$ width of $15472\pm{4274}$ km/s. But it can still be confirmed that Ly$\beta$ exhibits a strong broad component and is redshifted. During the re-observation of DESI in 2021 and SDSS in 2018, the raised bump was clearly below the detection limit. 

\vspace{-0.1cm}
\subsection{Physical Properties}

% Before measuring the accretion rate or Eddintong ratio, we need to obtain the bolometric luminosity and black hole mass for these four quasars. We determine the bolometric luminosity $L_{\rm bol}$ by applying a correction factor of $ L_{\rm bol}= 3.81 L_{\rm 1350}$ \citep{Richards2006,Shen2011}.  However, estimating the black hole mass at high redshift poses a significant challenge due to the unavailability of the size-luminosity ($R_{\rm BLR}$-$L_{5100}$) scaling relation for H$\beta$ from optical spectra. One effective method is to estimate the black hole mass by assuming the Keplerian motion of the gas that produces C {\sc iv} in a single-epoch spectrum, which is not particularly reliable since C {\sc iv} is typically associated with outflows \citep{Richards2011}. We calculated the black  hole mass using the formula proposed by \cite{Vestergaard2006}:

Before measuring the accretion rate or Eddington ratio, we determine these four quasars' bolometric luminosity and black hole mass. We estimate the bolometric luminosity, $L_{\rm bol}$, by applying the correction factor $ L_{\rm bol}= 3.81 L_{\rm 1350}$ \citep{Richards2006, Shen2011}. However, estimating black hole mass at high redshift is challenging. One common approach is to estimate black hole mass assuming the gas producing C\,{\sc iv} follows Keplerian motion, using a single-epoch spectrum. This method, however, is not particularly reliable since C\,{\sc iv} is often associated with outflows \citep{Richards2011}. We calculated the black hole mass using the formula from \cite{Vestergaard2006}:
\begin{eqnarray}
\rm \log(M_{\bullet}/M_{\odot}) && = 6.66
+2.9\log( \rm\frac{FWHM_{\rm C IV}}{10^{3}\rm km\ s^{-1}}) \nonumber\\ 
&&+0.53\log(\frac{\lambda L_{\lambda}(1350\rm \AA)}{10^{44} \rm erg \ s^{-1}}),
\end{eqnarray}
where $\rm FWHM_{\rm C\,IV}$ represents the velocity width of C\,{\sc iv}. The results, presented in Table \ref{tab_measure}, show that the black hole masses of these four sources reach up to $M_{\bullet}=10^{9}M_{\odot}$, which is measuring through the spectrum from bright state.

The Eddington ratio ($\lambda_{\rm Edd} = L_{\rm bol} / L_{\rm Edd}$) is also presented in Table \ref{tab_measure}, along with the accretion rate, 
$\mathscr{\dot M}\approx \eta \lambda_{\rm Edd}$, assuming  $\eta=0.1$. The Eddington ratios of the four objects are notably low ($\lambda_{\rm Edd} = 0.072 \sim 0.004$), especially for spectra in the low states, which aligns with predictions of the Advection-Dominated Accretion Flow (ADAF) model ($\mathscr{\dot M}<10^{-3}\sim 10^{-2}$, \citealt{Narayan1994, Yuan2001, Wang2012}). These findings of low accretion rates are consistent with previous studies on CL AGN or quasars \citep{MacLeod2016, Green2022}. The general consistency of CL events between UV and optical broad emission lines suggests that changes in accretion primarily drive these CL events.

We hypothesize that the CL events of UV emission lines are also associated with a state transition in the accretion disk (between a canonical thin disk and the ADAF regime). Despite our luminosity measurements being significantly higher than the BLR disappearance threshold predicted by \cite{Elitzur2009},  the  Ly$\alpha$ and C\,{\sc iv} CL events,  produced in the inner area of the BLR and closest to the accretion disk,  would be an indicator of AGN activity change. A similar scenario is seen in Mrk 590, where a decrease in continuum luminosity, accompanied by a CL event in the H$\beta$ emission line in the optical band, leads to significant flux variations in Ly$\alpha$ and C\,{\sc iv}, which almost disappear, leaving only faint and weak broad components in high-SNR UV spectra \citep{denney2014, Mathur2018}. The physical mechanism behind the CL events in these four quasars or candidates might be similar to that of Mrk 590, mainly driven by  accretion changes.

\subsection{Unexpected CL Relationship between  Ly$\alpha$ and C {\sc iv}}

\cite{Ross2020} and \cite{Guo2020} reported several C\,{\sc iv} CL AGNs/quasars where a significant Ly$\alpha$ broad component was always present. \cite{Guo2020} suggested a CL sequence (Mg\,{\sc ii}, C \, {\sc iv} , Ly$\alpha$) in which Ly$\alpha$ might be the last broad emission line to disappear after the broad C\,{\sc iv} vanishes, or the first broad emission line to appear before the broad C\,{\sc iv} emerges in the spectrum. 
\cite{Guo2020} suggest that the transition sequence may be influenced by the decay rates of different emission lines. They adopt the locally optimally emitting cloud  model to simulate the sequence under a decreasing continuum and find that the decay rates vary with diminishing luminosity (see details in Figure 10). We initially expected the CL event of C\,{\sc iv} to occur after the Ly$\alpha$ event in these four quasars. However, as shown in Figure \ref{fig_1}, although the increase (or decrease) in C\,{\sc iv} intensity is accompanied by a corresponding change in continuum luminosity, consistent with previous studies on C\,{\sc iv} variation \citep{Wilhite2006}, the CL sequence between broad C\,{\sc iv} and Ly$\alpha$ is unexpected from the simulation of \cite{Guo2020}. The C\,{\sc iv} emission lines of J1512 and J1306 exhibit significant variations, though with smaller changes in FWHM compared to the broad Ly$\alpha$ line. In contrast, the C\,{\sc iv} emission lines of J1602 and J1511 are near the detection limit or nearly disappear in their low states.

When combined with the Ly$\alpha$ and C\,{\sc iv} CL quasars reported by \cite{Ross2020} and \cite{Guo2020}, it appears that there is no precise sequence of CL events between Ly$\alpha$ and C\,{\sc iv}, complicating the physical explanation of CL phenomena. In most cases, we expect the CL events in Ly$\alpha$ and C {\sc iv} to occur on similar timescales, particularly since both emission lines originate from the highly ionized region of the quasar. However, the observation that the broad component of C {\sc iv} can still persist after Ly$\alpha$ undergoes a CL event suggests that changes in these emission lines are not always synchronized and may be influenced by different physical mechanisms. This phenomenon could be related to the emission regions and dynamical properties of Ly$\alpha$ and C {\sc iv}. For example, C {\sc iv} is typically associated with gas outflows in more extended regions \citep{Richards2011, Sun2018}, while Ly$\alpha$ emission may arise closer to the black hole’s accretion disk. Additionally, associated with accretion change, partial obscuration or chaotic gas motion might also together result in C {\sc iv} remaining visible even after changes in Ly$\alpha$. The asynchrony presents additional challenges in understanding the physical origin of CL events and requires further multi-wavelength observations, especially in the X-ray and infrared bands, to detailedly analyze the changing conditions in the nuclear region \citep{denney2014, Yang2023}, rather than attributing CL solely to  the accretion rate change.

% good to comment on how these systems are different to NGC 5548 Mrk 509

\section{Summary}
\label{sec_conclusion}

This study reports two Ly$\alpha$ CL quasars and two candidates, identified through repeated observations from SDSS and DESI. We find that the Eddington ratios or accretion rates of the four targets fall between the Advection-Dominated Accretion Flow and canonical disk regimes ($\mathscr{\dot M} < 10^{-3} \sim 10^{-2}$). However, when combined with the findings of \cite{Ross2020}, the CL behavior of the two high-ionization C {\sc iv} and Ly$\alpha$ lines appears confusing, contradicting the CL sequence proposed by \cite{Guo2020}. The unexcepted CL sequence suggests that the physical properties and origins of the two broad emission lines may differ and be much more complex, warranting further observation and study. These rare CL events at high redshift provide insights into the response of highly ionized emission lines from the innermost BLR to changes in continuum luminosity, offering a new perspective on the evolution of high-redshift quasars and galaxies.

\section*{acknowledgements}

Wei-Jian Guo thank DESI internal review (Tamara Davis and David Alexander) for very helpful discussions and comments  and thank DESI PubBoard Handler (Benjamin Joachim) to provide timely help. 
The work is supported by the funding supports from the National Key R\&D Program of China (grant Nos. 2023YFA1607800 and 2022YFA1602902) and Strategic Priority Research Program of the Chinese Academy of Sciences with Grant Nos. XDB0550100 and XDB0550000. The authors also acknowledge the supports from the National Natural Science Foundation of China (NSFC; grant Nos. 12120101003, 12373010, 12173051, and 12233008), the National Key R\&D Program of China (2023YFA1607804, 2023YFA1608100, and 2023YFF0714800), Beijing Municipal Natural Science Foundation (grant No. 1222028), and China Manned Space Project with Nos. CMS-CSST-2021-A02, CMS-CSST-2021-A04 and CMS-CSST-2021-A05.  M.S. acknowledges support by the Polish National Agency for Academic Exchange (Bekker grant BPN/BEK/2021/1/00298/DEC/1), the State Research Agency of the Spanish Ministry of Science and Innovation under the grants 'Galaxy Evolution with Artificial Intelligence' (PGC2018-100852-A-I00) and 'BASALT' (PID2021-126838NB-I00). This work was partially supported by the European Union's Horizon 2020 Research and Innovation program under the Maria Sklodowska-Curie grant agreement (No. 754510).

This material is based upon work supported by the U.S. Department of Energy (DOE), Office of Science, Office of High-Energy Physics, under Contract No. DE–AC02–05CH11231, and by the National Energy Research Scientific Computing Center, a DOE Office of Science User Facility under the same contract. Additional support for DESI was provided by the U.S. National Science Foundation (NSF), Division of Astronomical Sciences under Contract No. AST-0950945 to the NSF’s National Optical-Infrared Astronomy Research Laboratory; the Science and Technology Facilities Council of the United Kingdom; the Gordon and Betty Moore Foundation; the Heising-Simons Foundation; the French Alternative Energies and Atomic Energy Commission (CEA); the National Council of Humanities, Science and Technology of Mexico (CONAHCYT); the Ministry of Science, Innovation and Universities of Spain (MICIU/AEI/10.13039/501100011033), and by the DESI Member Institutions: \url{https://www.desi.lbl.gov/collaborating-institutions}.

The DESI Legacy Imaging Surveys consist of three individual and complementary projects: the Dark Energy Camera Legacy Survey (DECaLS), the Beijing-Arizona Sky Survey (BASS), and the Mayall z-band Legacy Survey (MzLS). DECaLS, BASS and MzLS together include data obtained, respectively, at the Blanco telescope, Cerro Tololo Inter-American Observatory, NSF’s NOIRLab; the Bok telescope, Steward Observatory, University of Arizona; and the Mayall telescope, Kitt Peak National Observatory, NOIRLab. NOIRLab is operated by the Association of Universities for Research in Astronomy (AURA) under a cooperative agreement with the National Science Foundation. Pipeline processing and analyses of the data were supported by NOIRLab and the Lawrence Berkeley National Laboratory. Legacy Surveys also uses data products from the Near-Earth Object Wide-field Infrared Survey Explorer (NEOWISE), a project of the Jet Propulsion Laboratory/California Institute of Technology, funded by the National Aeronautics and Space Administration. Legacy Surveys was supported by: the Director, Office of Science, Office of High Energy Physics of the U.S. Department of Energy; the National Energy Research Scientific Computing Center, a DOE Office of Science User Facility; the U.S. National Science Foundation, Division of Astronomical Sciences; the National Astronomical Observatories of China, the Chinese Academy of Sciences and the Chinese National Natural Science Foundation. LBNL is managed by the Regents of the University of California under contract to the U.S. Department of Energy. The complete acknowledgments can be found at \url{https://www.legacysurvey.org/}.

Any opinions, findings, and conclusions or recommendations expressed in this material are those of the author(s) and do not necessarily reflect the views of the U. S. National Science Foundation, the U. S. Department of Energy, or any of the listed funding agencies.

The authors are honored to be permitted to conduct scientific research on Iolkam Du’ag (Kitt Peak), a mountain with particular significance to the Tohono O’odham Nation.

SDSS is managed by the Astrophysical Research Consortium for the Participating Institutions of the SDSS Collaboration including the Brazilian Participation Group, the Carnegie Institution for Science, Carnegie Mellon University, Center for Astrophysics | Harvard \& Smithsonian, the Chilean Participation Group, the French Participation Group, Instituto de Astrofísica de Canarias, The Johns Hopkins University, Kavli Institute for the Physics and Mathematics of the Universe(IPMU)/University of Tokyo, the Korean Participation Group, Lawrence Berkeley National Laboratory, Leibniz Institut fürAstrophysik Potsdam (AIP), Max-Planck-Institut für Astronomie (MPIA Heidelberg), Max-Planck-Institut für Astrophysik (MPA Garching), Max-Planck-Institut für ExtraterrestrischePhysik (MPE), National Astronomical Observatories of China, New Mexico State University, New York University, University of Notre Dame, Observatário Nacional/MCTI, The Ohio State University, Pennsylvania State University, Shanghai Astronomical Observatory, United Kingdom Participation Group, Universidad Nacional Autónoma de México, University of Arizona, University of Colorado Boulder, University of Oxford, University of Portsmouth, University of Utah, University of Virginia, University of Washington, University of Wisconsin, Vanderbilt University, and Yale University.

\appendix
\section{spectral decomposition}
\label{appendix}

To derive the continuum luminosity and emission line components, we utilize the \texttt{DASpec} software\footnote{\url{https://github.com/PuDu-Astro/DASpec}} to decompose the spectrum, following the method described in \cite{Kriss2019} for modeling Ly$\alpha$ in NGC 5548. Specifically, one Gaussian is used to model the narrow component and two Gaussians for the broad components in the bright-state spectrum. We adopt a single Gaussian for the dim state to fit the narrow component. Since all high-redshift quasars are sufficiently bright, resulting in negligible starlight contamination, the decomposition excludes any contribution from the host galaxy. We assume non-shifted Gaussians to model the broad Ly$\alpha$ because its blue side is absorbed by intergalactic material (IGM).

\begin{figure*}[t!]
\centering
\vspace{-0.8cm}
\includegraphics[width=0.8\textwidth]{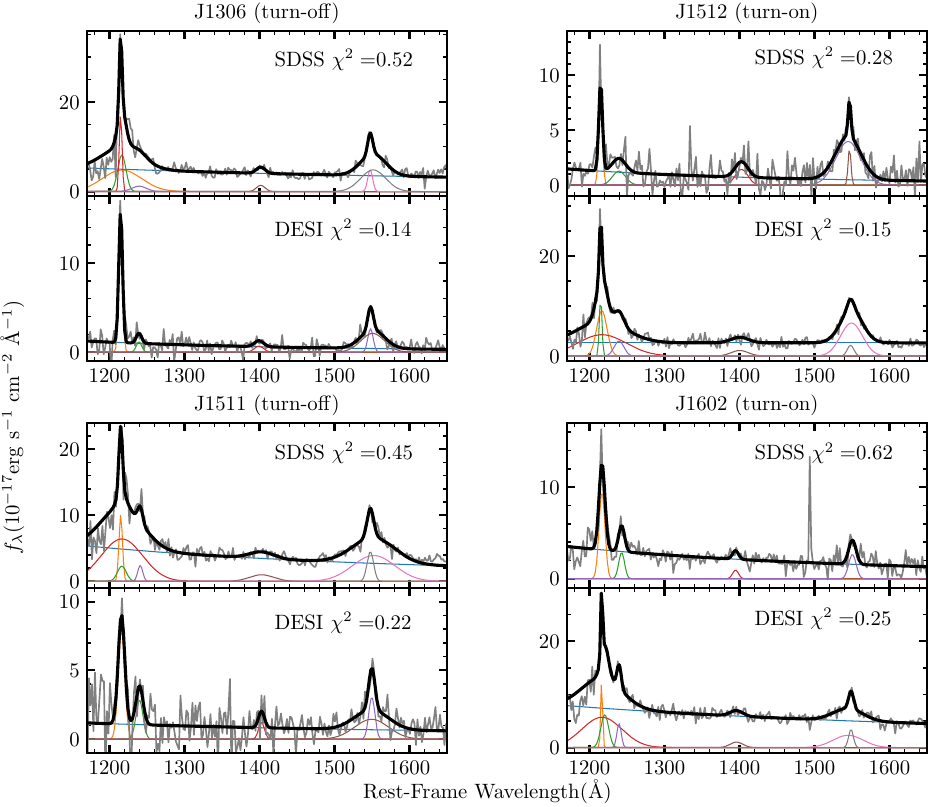}
\caption{ The spectral decomposition  of J1306,  J1512, J1511 and J1602. The grey lines are original spectra and the black lines are best-fitting results.  The blue lines represent the continuum. The colored lines repressents the narrow emission lines and broad emission lines.}
\label{fig_2}
\end{figure*}

The applied model components are 1) a power-law with no spectral features at 1350~\AA; 2) a single Gaussian for each broad emission line of N {\sc v} and S {\sc iv}; 3) two Gaussians, one narrow and one broad, for C {\sc iv}; 4) three Gaussians for Ly$\alpha$ in the bright state or one Gaussian for Ly$\alpha$ in the dim state. However, as shown in Figure \ref{fig_2}, the spectral decomposition may not be robust for the four targets due to the low signal-to-noise ratio (SNR).

For J1511 and J1602, the full width at half maximum (FWHM) of Ly$\alpha$ is between 1000 km/s and 2000 km/s. Although this width is slightly larger than that of a typical narrow emission line, we model the Ly$\alpha$ with a single Gaussian for two reasons: 1) the black hole masses of J1511 and J1602 are $\log (M_{\bullet}/M_{\odot}) \geq 9.0$, which causes the narrow emission lines to broaden; 2) modeling Ly$\alpha$ with two Gaussians results in a broad component with FWHM $\leq$ 4000 km/s, which is inconsistent with typical CL-AGN, where broad emission lines in the dim state are expected to be much broader than in the bright state. Therefore, the two-Gaussian model is less appropriate for J1511 and J1602.

\bibliographystyle{aasjournal}
\bibliography{sample.bib}

\end{document}